\documentclass[aps,twocolumn,floats,superscriptaddress,prd,nofootinbib,preprintnumbers]{revtex4-1}
\usepackage{graphicx, epsfig, bm, amsmath}

\usepackage{color}
\usepackage{hyperref}
\usepackage{wasysym}
\usepackage{tensor}
\usepackage{myterms}

\begin{document}


\newcommand{\lcdm}{$\Lambda$CDM}
\newcommand{\gpr}{G^{\prime}}
\newcommand{\fnl}{f_{\rm NL}}
\newcommand{\curv}{{\cal R}}
\def\nn{\nonumber}
\def\({\left(}
\def\){\right)}
\def\[{\left[}
\def\]{\right]}
\def\tr{{\rm Tr}}
\def\gfc{\xi}
\newcommand{\beq}{\begin{equation}}
\newcommand{\beqn}{\begin{eqnarray}}
\newcommand{\eeq}{\end{equation}}
\newcommand{\eeqn}{\end{eqnarray}}
\newcommand{\GW}{{\rm GW}}
\newcommand{\RH}{\text{\sc rh}}
\definecolor{darkgreen}{cmyk}{0.85,0.2,1.00,0.2}
\newcommand{\gB}{g_B}
\newcommand{\WP}{W}
\newcommand{\XP}{X}
\newcommand{\B}{B^{\rm Bulk}}
\renewcommand{\L}{\textsf{L} }
\renewcommand{\B}{\textsf{B} }
\newcommand{\CP}{\textsf{CP}}
\newcommand{\aap}{Astron. Astrophys.}


\pagestyle{plain}

\title{Gravitational Leptogenesis, Reheating, and Models of Neutrino Mass}

\author{Peter Adshead}
\affiliation{Department of Physics, University of Illinois at Urbana-Champaign, Urbana, Illinois 61801, U.S.A.}

\author{Andrew J. Long}
\affiliation{Kavli Institute for Cosmological Physics, University of Chicago, Chicago, Illinois 60637, U.S.A.}
      
\author{Evangelos I. Sfakianakis}
\affiliation{Department of Physics, University of Illinois at Urbana-Champaign, Urbana, Illinois 61801, U.S.A.}
\affiliation{Nikhef, Science Park 105, 1098 XG Amsterdam, The Netherlands}
\affiliation{Institute Lorentz of Theoretical Physics, University of Leiden, 2333CA Leiden, The Netherlands}

\preprint{Nikhef 2017-50}

\begin{abstract}
Gravitational leptogenesis {refers to a class of baryogenesis models} in which the matter-antimatter asymmetry of the universe arises through the standard model lepton-number gravitational anomaly.  
In these models chiral gravitational waves source a lepton asymmetry in standard model neutrinos during the inflationary epoch.  
We point out that gravitational leptogenesis can be successful in either the Dirac or Majorana neutrino mass scenario.  
In the Dirac mass scenario, gravitational leptogenesis predicts a relic abundance of sterile neutrinos that remain out of equilibrium, and the lepton asymmetry carried by the standard model sector is unchanged.  
In the Majorana mass scenario, the neutrinos participate in lepton-number-violating interactions that threaten to washout the lepton asymmetry during post-inflationary reheating.  
However, we show that a complete (exponential) washout of the lepton asymmetry is prevented if the lepton-number-violating interactions go out of equilibrium before all of the standard model Yukawa interactions come into equilibrium.  The baryon and lepton asymmetries carried by right-chiral quarks and leptons are sequestered from the lepton-number violation, and the washout processes only suppress the predicted baryon asymmetry by a factor of $\varepsilon_{\rm w.o.} = \pm O(0.1)$.  
The sign of $\varepsilon_{\rm w.o.}$ depends on the model parameters in such a way that a future measurement of the primordial gravitational wave chirality would constrain the scale of lepton-number violation (heavy Majorana neutrino mass).  
 
\end{abstract}

\maketitle  

\section{Introduction}
\label{sec:intro}

Our observable Universe is overwhelmingly dominated by matter, rather than antimatter.  This asymmetry is quantified by the dimensionless ratio $n_\B / s$, where $n_\B$ is the number density of baryon number and $s$ is the entropy density of the cosmological plasma.  The baryon relic abundance is measured from observations of the cosmic microwave background (CMB) to be $\Omega_b h^2 \simeq (0.0223 \pm 0.0002)$ \cite{Planck:2015xua}, which implies a baryon-to-entropy ratio of 
\begin{align}\label{eq:BAU_measure}
Y_\B \equiv \frac{n_\B}{s} \simeq \(0.861 \pm 0.008\)\times 10^{-10} \per
\end{align}
Observations of the light element abundances furnish a consistent measurement of $Y_\B$ when compared with the predictions of big bang nucleosynthesis (see, e.g., \cite{Fields:2014uja}).

The origin of this small asymmetry has long been a mystery. Inflation dilutes the number density of any pre-existing relics by a factor of $e^{-3N}$, where $N \sim 50$, and implies that any matter-antimatter asymmetry must be generated during the subsequent evolution of the Universe. Sakharov long ago enumerated the conditions for the successful dynamical generation of the baryon asymmetry \cite{Sakharov:1967dj}, and subsequently many models for baryogenesis have been proposed. 

Leptogenesis models \cite{Strumia:2006qk} generate the asymmetry first in the lepton sector (see, for example, \cite{Chen:2007fv}) and then invoke the electroweak sphaleron process to distribute the asymmetry between the leptons and the baryons. Several of these models employ inflationary or immediate post-inflationary dynamics to produce the lepton asymmetry (including but not limited to \cite{Adshead:2015jza, Yang:2015ida, Pearce:2015nga}). In these models, the lepton asymmetry is usually first manifested in the neutrino sector.

Multiple observations of neutrino flavor oscillations have now established that at least two of the neutrino species have non-zero masses \cite{Agashe:2014kda, Wang:2015rma}. Whereas massless fermions are uniquely described by Weyl spinor fields, massive fermions can be described by either Majorana or Dirac spinors depending on whether the particles are self-conjugate under charge conjugation, $\mathsf{C}$.  At present, the particle nature of the neutrinos (Dirac or Majorana) remains an open question. In fact, neutrinos may be the first elementary Majorana fermions known to us \cite{Kayser:2009zz}.  

The nature of the neutrinos (Dirac or Majorana) is crucial for many models of inflationary leptogenesis that produce the lepton asymmetry initially in the neutrino sector.  If the neutrinos are Dirac fermions, then equal and opposite lepton number is produced in the left-handed standard model (SM) {neutrinos} and their right-handed sterile partners, and no net lepton-number asymmetry arises.   If neutrinos are Majorana fermions instead, then  lepton-number-violating interactions can partly (or even completely) washout the resulting asymmetry. This is similar to washout processes that are known to occur in models of thermal leptogenesis \cite{Giudice:2003jh}. 

In this paper we point out that gravitational leptogenesis is compatible with either neutrino mass scenario. We also point out that {for the Majorana scenario}, gravitational leptogenesis \emph{does not} require the scale of lepton-number violation (e.g., mass scale of heavy Majorana neutrinos) to satisfy $m_{N} \gg H_{I}$, where $H_{I}$ is the Hubble scale during inflation.  We further estimate the effect of lepton-number-violating processes that washout the baryon asymmetries on the parameter space of generic inflationary gravitational leptogenesis scenarios in the Majorana scenario.

\newpage 
This paper is organized as follows. In Section \ref{sec:GravLepto}, we review the basic mechanism of gravitational leptogenesis and discuss its various implementations.  In Section \ref{sec:BaryonAsym} we generalize the assumption of instantaneous reheating and compute how the baryon asymmetry is diluted during the epoch of reheating.  Up to this point we assume that baryon-minus-lepton number is conserved, as in the standard model, and in Section \ref{sec:neutrinomasses} we discuss how the Dirac and Majorana neutrino mass scenarios affect gravitational leptogenesis.  In Section \ref{sec:washout}, we explore the Majorana mass scenario more carefully and calculate the predicted baryon asymmetry for gravitational leptogenesis.  We summarize our results in Section \ref{sec:conclusions}. Throughout we work in natural units where $\hbar = c = k_B = 1$, and we explicitly retain the reduced Planck mass $M_{\rm Pl} = (8\pi G_N)^{-1/2}$.  

\section{Gravitational leptogenesis}\label{sec:GravLepto}

In the standard model of particle physics, baryon number ($\B$) and lepton number ($\L$) are not conserved charges, but rather the corresponding symmetries, $\U{1}_\B$ and $\U{1}_\L$, are violated by quantum effects.  Specifically, $\L$ develops a gravitational anomaly, because gravity couples to left-chiral neutrinos that have no right-chiral counterpart in the standard model \cite{Kumura:1969wj,Delbourgo:1972xb,Eguchi:1976db,Gibbons:1979kq}, but there is no gravitational anomaly for $\B$ since the standard model contains equal numbers of left- and right-chiral quarks.  It is natural to ask whether the lepton-number gravitational anomaly can be used to explain the observed matter-antimatter asymmetry of the universe \cite{Ibanez:1992hb}.  Gravitational leptogenesis \cite{Alexander:2004us} is an elegant implementation of that idea.  

Gravitational leptogenesis refers to a class of models in which chiral gravitational waves are generated during the inflationary epoch, and the resulting non-zero gravitational Pontryagin density sources a lepton asymmetry.  This is quantified by the current conservation equation \cite{AlvarezGaume:1983ig}\footnote{Considering quantum electrodynamics with a single flavor of vector-like fermions, Refs.~\cite{Kumura:1969wj,Eguchi:1976db,Gibbons:1979kq} derive $\partial_\mu J^\mu_A = (-1/12)(1/16\pi^2) R\tilde{R}$ for the anomalous divergence of the axial vector current.  The vector current is exactly conserved, $\partial_\mu J^\mu_V = 0$.  Consequently, the chiral currents $J_{L,R} = (J_V \mp J_A) / 2$ obey $\partial_\mu J^\mu_{L,R} = (\pm1/24)(1/16\pi^2) R\tilde{R}$.  (The calculation of $\partial J_A$ in \rref{Delbourgo:1972xb} contains a factor of $2$ error, and the calculation in \rref{AlvarezGaume:1983ig} differs by a factor of $2$ because they consider chiral fermions for which $\partial J_R = 0$.) The standard model lepton-number current is $J_\L = \sum_i J_{e_L^i} + J_{e_R^i} + J_{\nu_L^i}$, where the index is summed over three generations.  The electrons have vector-like gravitational interactions, and their contributions to $\partial_\mu J_\L^\mu$ cancel leaving only the contribution from the three left-chiral neutrinos.  }
\begin{align}\label{eq:anom_eqn}
	{\partial_\mu \bigl( \sqrt{-g} J_{\B-\L}^\mu \bigr) = -\frac{N_{L-R} }{24} \frac{1}{16\pi^2} \, R\tilde{R} \per}
\end{align} 
The coefficient is $N_{L-R} = -\sum_i \chi_i (\B_i - \L_i)$, which sums all the Weyl spinor fields in the theory counting $\chi_i = +1$ ($-1$) for each left-chiral (right-chiral) spinor and weighting the sum by the baryon-minus-lepton number of each field ($\B_i-\L_i$). In the standard model $N_{L-R} = 3$, and a growing gravitational wave chirality therefore sources net lepton number in the form of a net left-handed neutrino  asymmetry.  

Most studies of gravitational leptogenesis assume either that the neutrinos are massless, as in the standard model, or that they are Majorana particles, and the scale of lepton-number violation is much higher than the energy scale of inflation so that the new degrees of freedom can be neglected.  In Section \ref{sec:neutrinomasses} we discuss the effect of finite neutrino mass on models of gravitational leptogenesis, and in Section \ref{sec:washout} we show how the predicted baryon asymmetry (including its sign) depends on the details of the neutrino mass generation. 

While the basic mechanism of gravitational leptogenesis from chiral gravitational wave production during inflation is robust, the original model proposed in Ref.\ \cite{Alexander:2004us} has a number of issues. In this scenario, chiral gravitational waves are generated via the coupling of a pseudo-scalar inflaton to the gravitational Chern-Simons term, or Pontryagin density \cite{Lue:1998mq, Jackiw:2003pm} (see also \rref{Kawai:2017kqt}). However, it has been argued that this coupling makes the predictions of the theory sensitive to unknown ultraviolet (UV) physics \cite{Lyth:2005jf}. Further, in this realization the majority of the contribution to the lepton current was argued to arise from graviton modes deep within the horizon. This results in an enhancement of the asymmetry by a factor of $(\Lambda / H_I)^4$, where $\Lambda$ is the ultraviolet cut-off scale of the theory and $H_I$ is the Hubble scale during inflation. In \rref{Alexander:2004us}, $\Lambda$ is taken to be the Planck scale. Ref.\  \cite{Fischler:2007tj} argues that once a proper renormalization procedure is applied, then this enhancement factor is removed, or effectively $\Lambda \sim H_I$.

A number of inflationary scenarios have been subsequently proposed in which large amplitude, chiral gravitational waves are abundantly produced in the absence of direct interactions between the inflaton and the gravitational Chern-Simons term. In the context of Natural Inflation \cite{Freese:1990rb}, a Chern-Simons interaction between the pseudo-scalar inflaton and a $\U{1}$ gauge field leads to the exponential production of helically-polarized gauge bosons \cite{Anber:2009ua}. These helical gauge bosons in turn generate a helically polarized gravitational wave spectrum \cite{Sorbo:2011rz}. Unfortunately, it has been recently shown that this mechanism does not generate a sufficient lepton asymmetry without spoiling inflation \cite{Papageorgiou:2017yup}.  Other more promising examples are inflationary scenarios that contain $\SU{2}$ gauge fields with classical vacuum expectation values, such as Gauge-flation \cite{Maleknejad:2011jw, Maleknejad:2011sq, Namba:2013kia} and its variants \cite{Nieto:2016gnp, Adshead:2017hnc}, Chromo-Natural Inflation (CNI) \cite{Adshead:2012kp,Dimastrogiovanni:2012ew,  Adshead:2013qp, Adshead:2013nka} and its variants \cite{Adshead:2016omu, Obata:2016tmo, Caldwell:2017chz}, and models that include spectator Chromo-Natural-like sectors \cite{Maleknejad:2016qjz, Dimastrogiovanni:2016fuu, Fujita:2017jwq, Agrawal:2017awz}.  Gravitational leptogenesis has been studied within the context of these $\SU{2}$ models \cite{Noorbala:2012fh, Maleknejad:2014wsa, Maleknejad:2016dci}; however, these works focused on the UV modes and suffer from similar criticisms regarding regularization and renormalization as the original proposal. 

More recently, Caldwell and Devulder pointed out that in a variant of CNI, large-amplitude chiral gravitational waves that leave the horizon near the end of inflation  could be responsible for the baryon asymmetry of the universe \cite{Caldwell:2017chz}. Furthermore, they demonstrated that requiring their model to generate a sufficient baryon asymmetry puts a lower bound on the tensor-to-scalar ratio that is accessible with upcoming Stage-4 CMB experiments \cite{Abazajian:2016yjj}. 

In the following we do not assume any specific implementation of gravitational leptogenesis, but we do use the work of \rref{Caldwell:2017chz} as a benchmark point for numerical estimates.

\section{Generation of Baryon Asymmetry}\label{sec:BaryonAsym}

In this section we calculate the baryon asymmetry that is generated through models of gravitational leptogenesis.  We treat the neutrinos as massless as predicted by the standard model, leaving the discussion of the issue of neutrino mass to Sections \ref{sec:neutrinomasses}~and~\ref{sec:washout}.  

Let $n_{\B-\L}(a)$ denote the number density of baryon number minus lepton number at a time when the FRW scale factor equals $a$, and thus $a^3 n_{\B-\L}$ is the comoving density.  During inflation $R\tilde{R} \neq 0$ causes {$a^3 | n_{\B-\L} |$} to grow, but after the end of inflation $R\tilde{R} = 0$ and $a^3 n_{\B-\L}$ is constant.  At this point we are ignoring the possibility of $(\B-\L)$-violating washout.  Furthermore, taking $R\tilde{R} = 0$ immediately after inflation amounts to neglecting the possibility of helical gravitational wave production during (p)reheating.

Let $a_e$ and $H_e$ denote the scale factor and Hubble parameter at the end of inflation.  It is convenient to introduce the dimensionless variable {$\Ncal_{\B-\L} = a^3 n_{\B-\L} / a_e^3 H_e^3$}, which represents the comoving number density of baryon-minus-lepton number per comoving Hubble volume at the end of inflation.  With the above assumptions, {$\Ncal_{\B-\L}$} is {constant} after the end of inflation ($a_e < a$).  

The lepton asymmetry produced from gravitational leptogenesis was first calculated in \rref{Alexander:2004us} (see also \rref{Caldwell:2017chz}).  The anomaly equation, Eq.\ \pref{eq:anom_eqn}, can be directly integrated by making use of the fact that $R\tilde{R} =  2\nabla_{\mu}K^{\mu}$, where $K^\mu$ is the topological current
\begin{align}
K^{\mu} = 2\epsilon^{\mu\alpha\beta\gamma}\left[\frac{1}{2}\Gamma^{\sigma}_{\alpha\tau}\partial_{\beta}\Gamma^{\tau}_{\gamma\sigma} + \frac{1}{3}\Gamma^{\sigma}_{\alpha\tau}\Gamma^{\tau}_{\beta\eta}\Gamma^{\eta}_{\gamma\sigma}\right], 
\end{align}
and $\Gamma$ is the usual Christoffel connection \cite{Jackiw:2003pm}. 
This leads to the change in the baryon-minus-lepton number during inflation (assuming an initially vanishing asymmetry at $t = t_i$)
\begin{align}
\label{eq:NL}
{\Ncal_{\B-\L}(t_e) = -2} \frac{3}{24} {1\over 16\pi^2} \left( {H_e \over M_{\rm Pl} } \right)^2 \Bigl( \Hcal_{R-L}^\GW(t_e)-\Hcal_{R-L}^\GW(t_i) \Bigr) \, .
\end{align}
The dimensionless quantity $\Hcal_{R-L}^\GW$ is the expectation value of the topological charge per unit Hubble volume at the end of inflation measured in units of the standard gravitational wave power spectrum amplitude \cite{Caldwell:2017chz}
\begin{align}
\label{eq:H_integral_form}
	\Hcal_{R-L}^\GW \equiv  \int \! \ud \ln k \, \left [  {k^3\over H_e^3 } {( \Delta_R^2 - \Delta_L^2 )\over H_e^2/M_{\rm Pl}^2} - {k\over H_e}  {(\Delta_R^{\prime 2} - \Delta_L^{\prime 2})  \over  H_e^4/M_{\rm Pl}^2} \right ] .
\end{align}
Here $\Delta_\chi^2(k,\tau) = (k^3/2\pi^2) |\gamma_\chi(k ,\tau) |^2$ is the dimensionless power spectrum for gravitational waves of chirality $\chi \in \{ L,R \}$, $\Delta_\chi^{\prime 2} = (k^3/2\pi^2)|\partial_\tau\gamma_\chi(k ,\tau) |^2 $, $\gamma_\chi$ are the amplitudes of the left- and right-helicity gravitational waves, and $k$ is the comoving wavenumber.  

Although $\Hcal_{R-L}^\GW$ is model-dependent, we can still make a few general comments on its properties. In order to produce any significant particle asymmetry per Hubble volume ($\Ncal_{\B-\L} > 1$), we clearly require $\Hcal_{R-L}^\GW \gg 1$, because CMB constraints impose $(H_e/M_{\rm Pl}) \lesssim 10^{-5}$.  Examining Eq.\ \eqref{eq:H_integral_form} suggests two ways of obtaining a large $\Hcal_{R-L}^\GW$:
\begin{enumerate}
\item  A spectrum of chiral gravitational waves of the typical inflationary amplitude ($\Delta_R^2-\Delta_L^2\sim \mathcal{O}(\epsilon_H )({H_e / M_{\rm Pl} })^2$) that contributes up to some far UV scale $\Lambda \gg H_e$, (see, for example, \cite{Alexander:2004us}). 
\item Gravitational wave modes of one helicity that attain a very large amplification -- above their usual inflationary values --  near to the horizon at the end of inflation  ($\Delta_R^2-\Delta_L^2 \gg ({H_e / M_{\rm Pl} })^2$ for $k\sim a_e H_e$) \cite{Caldwell:2017chz}.  
\end{enumerate}
In what follows, we remain agnostic about the origin of such a large topological charge, although we take as a benchmark the value we estimate from \rref{Caldwell:2017chz} of $\Hcal_{R-L}^\GW \sim -10^{14}$.

Let us now consider how the asymmetry in Eq.\ \pref{eq:NL} is distributed across the various standard model species.  For a left-chiral Weyl fermion $\chi$ let $n_\chi(t)$ be the number density of $\chi$-number at time $t$, i.e. the number density of left-handed $\chi$ particles minus the number density of right-handed $\bar{\chi}$ antiparticles.  Since the gravitational interaction is universal (flavor blind), gravitational leptogenesis produces an equal initial asymmetry in every standard model fermion species with only a differing sign for left- and right-chiral fermions.  The standard model fermions are denoted by $u_L^i$, $d_L^i$, $u_R^i$, $d_R^i$, $\nu_L^i$, $e_L^i$, and $e_R^i$, corresponding to the left-chiral up-type quarks of generation $i$ (color index suppressed), left-chiral down-type quarks, right-chiral up-type quarks, right-chiral down-type quarks, left-chiral neutral leptons, left-chiral charged leptons, and right-chiral charged leptons.  The number densities at the end of inflation satisfy 
\begin{subequations}\label{eq:initial}
\begin{align}
	n_{u_L^i}(t_e) = n_{d_L^i}(t_e) & = - N_c \, H_e^3 \, \Ncal_{\B-\L}(t_e) / 3\, , \\
	n_{u_R^i}(t_e) = n_{d_R^i}(t_e) & = + N_c \, H_e^3 \, \Ncal_{\B-\L}(t_e) / 3 \, ,\\
	n_{\nu_L^i}(t_e) = n_{e_L^i}(t_e) & = - H_e^3 \, \Ncal_{\B-\L}(t_e) / 3 \, ,\\ 
	n_{e_R^i}(t_e) & = + H_e^3 \, \Ncal_{\B-\L}(t_e) / 3\, ,
\end{align} 
\end{subequations}
where $i = 1,2,3$ is the generation index, and we have summed over $N_c = 3$ colors of quarks.  The initial asymmetries in the standard model bosons are zero.  Although the standard model quarks carry individual asymmetries, there is no initial baryon asymmetry, $n_\B = (1/3) \sum_i ( n_{u_L^i} + n_{d_L^i} + n_{u_R^i} + n_{d_R^i} ) = 0$.  The initial lepton asymmetry is $n_\L = \sum_i ( n_{\nu_L^i} + n_{e_L^i} + n_{e_R^i} ) = \sum_i n_{\nu_L^i} = - H_e^3 \, \Ncal_{\B-\L}$, which remains nonzero because the standard model neutrinos $\nu_L^i$ have no right-chiral counterpart.  This initial condition differs notably from thermal leptogenesis for which the initial asymmetry from heavy Majorana neutrino decays is carried only by the left-chiral leptons and the Higgs bosons.  

\begin{figure*}[t]
\begin{center}
\includegraphics[width=0.45\textwidth]{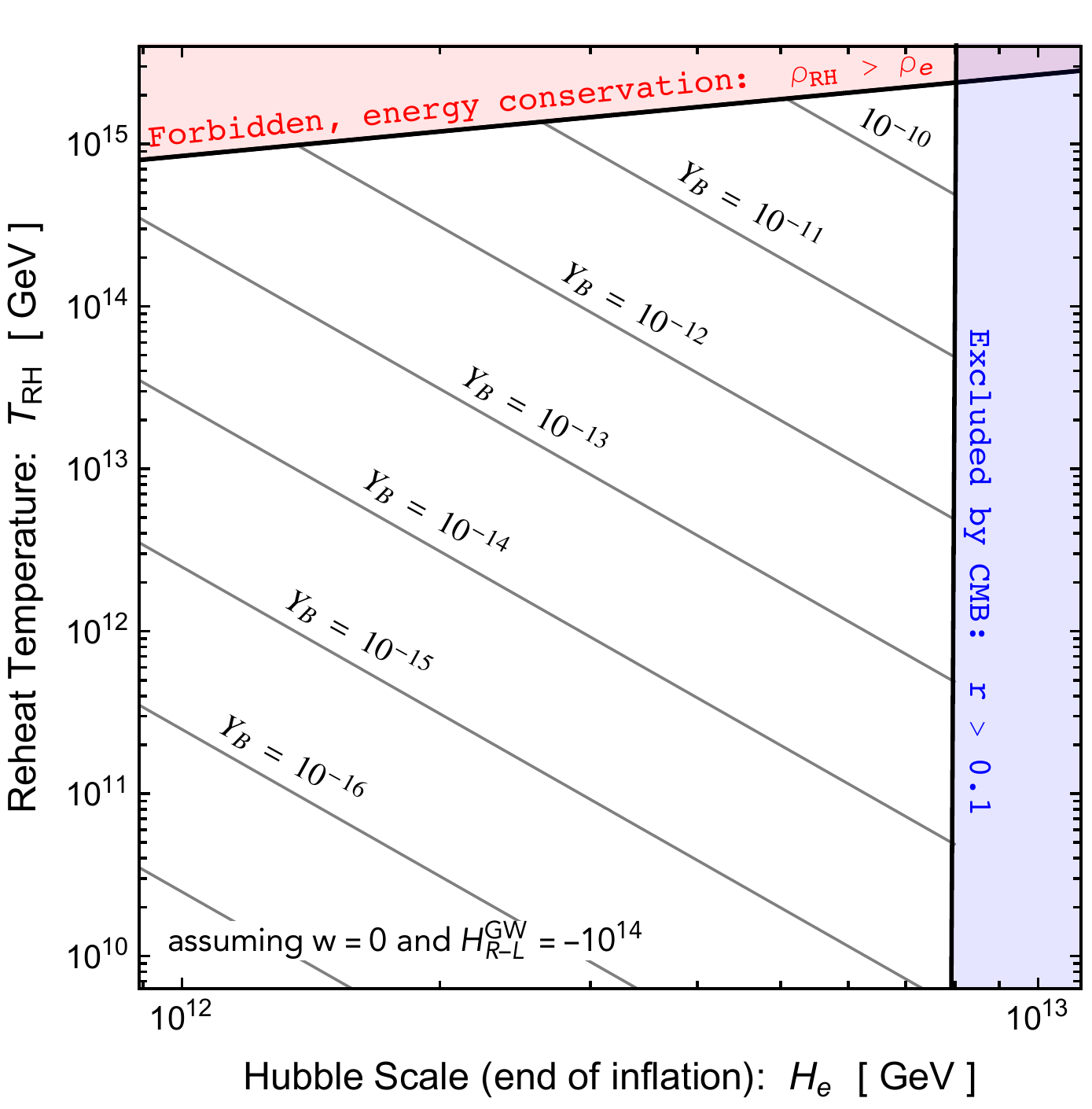} \hskip 1.5 cm \includegraphics[width=0.45\textwidth]{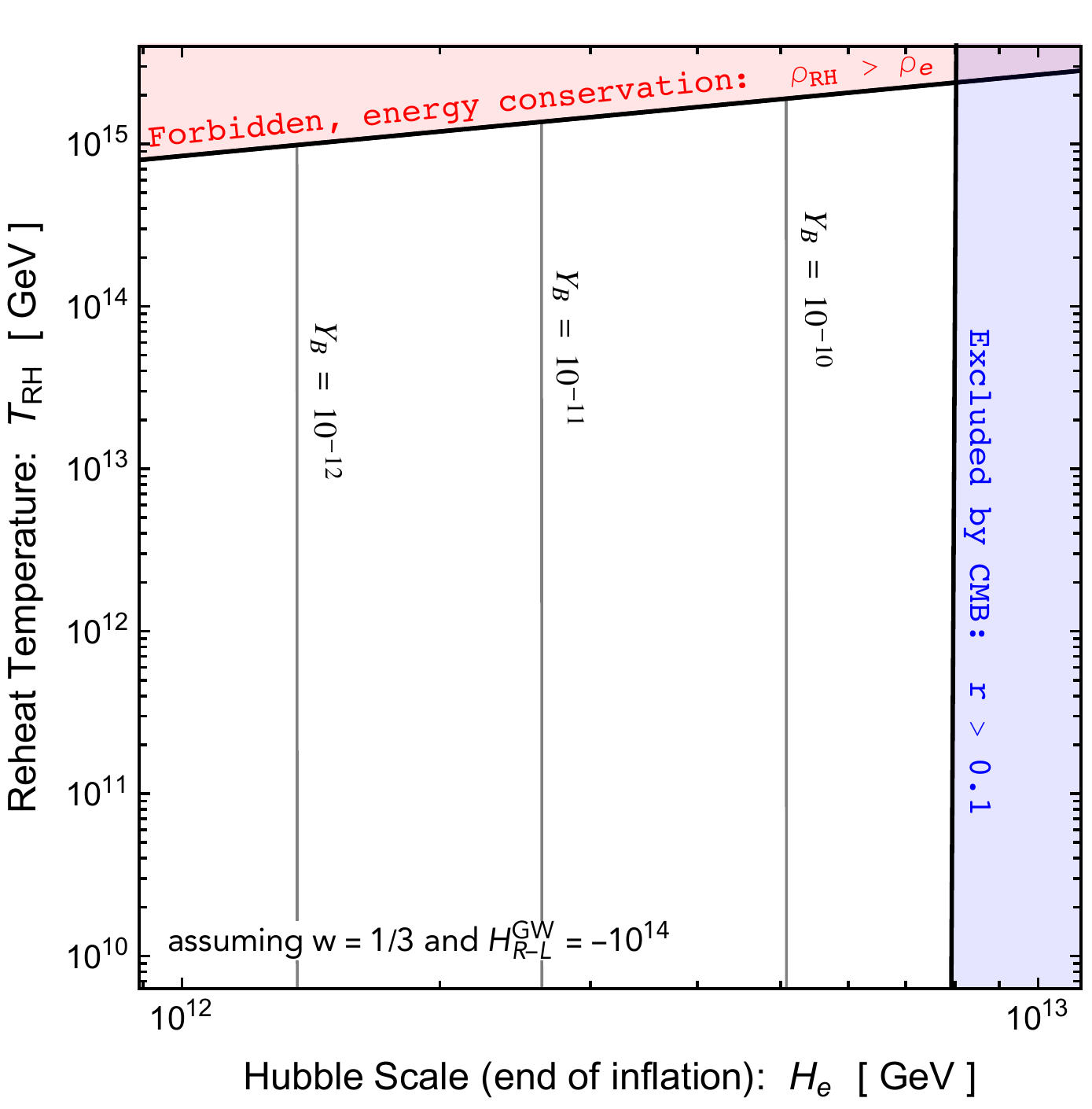}
\caption{\label{fig:param_space_1}
The baryon asymmetry $Y_\B = n_\B / s$ generated from gravitational leptogenesis in a model where the Hubble scale at the end of inflation is $H_e$, the plasma temperature at the end of reheating is $T_\RH$, and {baryon-minus-lepton}-number is assumed to be conserved after inflation (Dirac mass scenario). The left panel shows the case where the effective equation of state during reheating is $w = 0$ and the right panel shows $w = 1/3$.  {The amplitude of chiral gravitational waves is parametrized by $\Hcal_{L-R}^\GW$ [see \eref{eq:H_integral_form}], which} we take to be $\Hcal_{R-L}^\GW = -10^{14}$ in drawing the contours, but more generally $Y_\B \propto -\Hcal_{R-L}^\GW$ as in \eref{eq:etaB_numerical}. 
}
\end{center}
\end{figure*}

Next we calculate the baryon-minus-lepton asymmetry, $Y_{\B-\L} \equiv n_{\B-\L}/s$, which allows us to compare with the measured matter-antimatter asymmetry in Eq.\ \pref{eq:BAU_measure}.  Reheating occurs after the end of inflation ($a_e < a$) as the inflaton begins to transfer energy into relativistic particles. These relativistic particles thermalize quickly forming a plasma.  Eventually, the energy density of the plasma becomes larger than the energy density of the inflaton, which signals the end of reheating and the start of the radiation-dominated era.  We compute the  asymmetry after the inflaton has completely decayed and reheating is concluded, forming a thermal bath of SM particles.

Let $a_\RH$ denote the value of the  scale factor at the end of reheating.\footnote{We define the end of reheating to be the time when the Universe expands in a radiation dominated phase \emph{and} the SM is thermalized. The temperature at this time is denoted $T_{\RH}$. This is to allow for an equation of state $w = 1/3$ during reheating.}  Since the comoving number density {$\Ncal_{\B-\L}$} is (assumed to be) conserved after inflation, the physical number density of baryon-minus-lepton number at reheating is given by {$n_{\B-\L}(a_\RH) = (a_e/a_\RH)^3 H_e^3 \Ncal_{\B-\L}$}.  In general reheating has a finite duration ($a_e < a_\RH$), and the dilution factor $(a_e/a_\RH)^3$ measures the suppression of the asymmetry during this period.  For comparison, the entropy density of the plasma at this time is given by $s(a_\RH) = (2\pi^2/45) g_{\ast} T_\RH^3$ where $g_{\ast}$ is the effective number of relativistic species at temperature $T_\RH = T(a_\RH)$.  While the universe expands adiabatically {after the end of reheating}, the comoving entropy density $a^3 s$ is conserved, and therefore so too is the ratio {$Y_{\B-\L} = n_{\B-\L} / s$}.  

In order to evaluate the expansion factor $(a_e/a_\RH)^3$ we assume that the dominant energy component during reheating ($a_e < a < a_\RH$) can be described as a perfect fluid with pressure $p$, energy density $\rho$, and {\it constant} equation of state $w = p/\rho$.  
The continuity equation then yields
\begin{align}\label{eq:expansion_factor}
\frac{a_e}{a_\RH} = \(\frac{\rho_\RH}{\rho_e}\)^{\frac{1}{3(1+w)}} \per
\end{align}
The Friedmann equation gives $\rho_e = 3 M_{\rm Pl}^2 H_e^2$ where $\rho_e$ and $H_e$ are the cosmological energy density and the Hubble parameter at the end of inflation.  At the end of reheating, the energy density of the universe is dominated by the plasma, and the Friedmann equation gives $\rho_\RH = (\pi^2/30) g_{\ast} T_\RH^4$.  

Combining the equations above, we evaluate the baryon-minus-lepton asymmetry as 
\begin{align}\label{eq:etaL}
{Y_{\B-\L}} & = 2^{-\frac{2+w}{1+w}} 45^{\frac{w}{1+w}} \pi^{- \frac{2w}{1+w}} M_{\rm Pl}^{-\frac{2}{1+w}} 
\\ & \quad \times 
g_{\ast}^{\frac{-w}{1+w}} H_e^\frac{1+3w}{1+w} T_\RH^\frac{1-3w}{1+w} {\Ncal_{\B-\L}} \nonumber \per
\end{align}
If the effective equation of state during reheating is $w=0$ then ${Y_{\B-\L}} \sim (H_e T_\RH/M_{\rm Pl}^2)$, whereas if $w=1/3$ then ${Y_{\B-\L}} \sim (H_e/M_{\rm Pl})^{3/2}$, which is independent of $T_\RH$.\footnote{One could consider $w > 1/3$ following inflation, however, as noted by Ref.\ \cite{Papageorgiou:2017yup}, this generally requires a number of additional assumptions, and we do not consider it here. For $w <1/3$, ${Y_{\B-\L}}$ strictly decreases with the reheat temperature, and therefore instantaneous reheating represents an upper bound on the asymmetry attainable in this model.} Comparing the two cases, for a fixed $T_{\RH}$ and $H_e$, we see that a matter-dominated reheating stage produces a smaller baryon-minus-lepton asymmetry by a factor of roughly $T_\RH/\sqrt{H_e M_{\rm Pl}}$. This factor equals unity if $T_\RH$ is computed using the assumption of instantaneous reheating (the limiting case of all reheating scenarios) but can otherwise be very small if the matter-dominated stage is prolonged and the value of $T_\RH$ is significantly reduced.

In the standard model, both lepton number and baryon number are anomalous under the electroweak interactions \cite{Hooft:1976up}. Consequently, the lepton asymmetry is partially converted into a baryon asymmetry via non-perturbatively large thermal fluctuations of the $\SU{2}_L$ gauge field in the hot plasma (sometimes called the hot electroweak sphaleron) \cite{Kuzmin:1985mm,McLerran:1990de,Arnold:1996dy}.  If the initial baryon-minus-lepton asymmetry is given by ${Y_{\B-\L}}$ in \eref{eq:etaL}, then using the formalism of \rref{Harvey:1990qw} we calculate the final baryon asymmetry to be $Y_\B = (28/79) Y_{\B-\L}$.  

Using the above formulae, we evaluate the baryon asymmetry 
\begin{align}\label{eq:etaB_numerical}
Y_\B & \simeq \bigl( 4 \times 10^{-10} \bigr) \frac{C(w)}{C(0)} \left( \frac{g_{\ast}}{106.75} \right)^\frac{-w}{1+w} 
\\ & \quad \times 
\left( \frac{H_e}{10^{13} \GeV} \right)^{\frac{3+5w}{1+w} }\left( \frac{T_\RH}{10^{15} \GeV} \right)^\frac{1-3w}{1+w} \(\frac{ \Hcal_{R-L}^\GW }{-10^{14}}\) \nonumber 
\end{align}
where $C(w)$ is a numerical coefficient that can be inferred from \eref{eq:etaL}. For comparison, the observed value is $Y_\B \simeq 0.861 \times 10^{-10}$ from \eref{eq:BAU_measure}.  Therefore, gravitational leptogenesis is naturally accommodated in models with high-scale inflation ($H_e \gtrsim 10^{13} \GeV$), that produce large-amplitude, {left-}chiral gravitational waves at the end of inflation ($\Hcal_{R-L}^\GW \sim - 10^{14}$), provided reheating is efficient ($T_\RH \gtrsim 10^{15} \GeV$).  However, as we discuss in the next section, if the neutrinos are Majorana particles, this estimate is overly optimistic because washout effects have been neglected.  

We show the viable region of parameter space (neglecting washout) in \fref{fig:param_space_1} for reheating equations of state $w = 0$, as well as $w = 1/3$.  Energy conservation requires $\rho_\RH \leq \rho_e$, which implies an upper limit on the reheat temperature, $T_\RH \lesssim (3 \times 10^{15} \GeV) \sqrt{H_e / 10^{13} \GeV}$. Observations of the cosmic microwave background polarization (B-modes) impose an upper limit on the energy scale of inflation.  In models of single-field, slow-roll inflation, the amplitude of the tensor power spectrum is predicted to be $A_t = 2 H_{\rm cmb}^2 / (\pi^2 M_{\rm Pl}^2)$ where $H_{\rm cmb}$ is the value of the Hubble parameter when the modes that we observe today in the CMB were exiting the horizon during inflation, which is roughly 60 e-foldings before the end of inflation.  Planck measures the amplitude of the scalar power spectrum to be $A_s \simeq 10^{-10} e^{3.1}$, and it constrains the tensor-to-scalar ratio to be $r = A_t / A_s < 0.10$ \cite{Planck:2015xua}.  This implies an upper limit of $H_{\rm cmb} = \sqrt{(\pi^2/2) r A_s} M_{\rm Pl} \lesssim ( 8.0 \times 10^{13} \GeV) \sqrt{r / 0.1}$.  The relation between $H_{\rm cmb}$ and $H_e$ is model-dependent; for a quadratic inflaton potential we have $H_e \approx H_{\rm cmb} / 10$.  The next generation of CMB telescopes (Stage-3 and 4) are projected to be sensitive to $r$ at the level of $\sigma(r) \sim 0.01$ or better \cite{Abazajian:2016yjj}, which means that the entire parameter space in \fref{fig:param_space_1} can be tested with observations of CMB polarization \cite{Caldwell:2017chz}.  

\section{Implications of non-zero neutrino mass}\label{sec:neutrinomasses}

If the low energy particle content and interactions are described by the standard model, then gravitational leptogenesis works as we have described in the previous section.  However, the standard model must be extended in order to accommodate measurements of nonzero neutrino mass, which raises the question of whether the neutrinos are Dirac or Majorana particles.  In this section, we discuss each of these scenarios and their implications for gravitational leptogenesis.

\subsection{Massive Dirac neutrinos}
\label{sec:Dirac}

In the Dirac mass scenario, right-chiral neutrinos are added to the standard model and their mass is taken to be degenerate with the left-chiral neutrinos.  Consequently the gravitational anomaly in the lepton-number current is cancelled, {\it i.e.}, $N_{L-R} = 0$ in \eref{eq:anom_eqn}.  Nevertheless, gravitational leptogenesis is still viable.

Although a growing gravitational wave chirality does not generate a {\it net} lepton number, it does generate an axial-lepton number, {\it i.e.} equal and opposite lepton numbers in the left-chiral, active (SM) neutrinos and in the right-chiral, sterile neutrinos.  The conservation of axial-lepton number is violated by the neutrino Yukawa interaction, but since the Yukawa coupling is extremely small {($\lambda_\nu \sim m_\nu / v \simeq 10^{-12}$)}, these interactions are always out of equilibrium. Effectively, the lepton number carried by the right-chiral (sterile) neutrinos is sequestered from the baryon and lepton number carried by the standard model particles.  As a result, the asymmetries in the standard model sector are unaffected by the addition of the sterile neutrinos to the theory, and the outcome of gravitational leptogenesis is unchanged.  

This sequestration phenomenon is essentially a gravitational version of the well-known Dirac leptogenesis scenario \cite{Dick:1999je, Murayama:2002je}; in this model the initial axial-lepton number is generated through the gravitational anomaly instead of through the decay of a heavy species. We note that this scenario was not considered in the original gravitational leptogenesis proposal \cite{Alexander:2004us}. 

The sterile neutrinos persist today as a cosmological relic.  Their number density is approximately equal to the number density of baryon number, $n \simeq (3 \times 10^{-7}) \cm^{-3}$.  If these neutrinos are non-relativistic, then they contribute to the dark matter relic abundance.  Their energy density compared to the critical density is roughly $m n / (3 M_{\rm Pl}^2 H_0^2) \sim (6 \times 10^{-12}) (m / 0.1 \eV)$, which is a negligible contribution to the total dark matter relic abundance.  

\subsection{Massive Majorana neutrinos}\label{sec:majorana}

In the Majorana mass scenario, the neutrino masses arise from the lepton-number-violating Weinberg operator after electroweak symmetry breaking.\footnote{Here we assume that the scale of lepton-number violation, $m_N$, is much larger than the weak scale, $v$.  The regime $m_N \lesssim v$ starts to be constrained by experiment, but we note that $m_N \ll m_\nu$ is again unconstrained, and specifically the Dirac mass scenario is obtained in the limit $m_N \to 0$ (provided of course that the Yukawa couplings are changed appropriately to give the correct neutrino mass scale, $m_\nu$).}  The dimension-5 Weinberg operator can arise from various UV completions in which lepton number is violated.  One simple and compelling example is the (Type-I) seesaw model \cite{Minkowski:1977sc, Mohapatra:1979ia, GellMann:1980vs, Yanagida:1980xy, Mohapatra:1980yp, Schechter:1980gr}.  In this scenario, one introduces heavy right-handed Majorana neutrinos, and the Weinberg operator is generated upon integrating these particles out of the theory.  

Let us briefly anticipate the effect of massive Majorana neutrinos on gravitational leptogenesis; we postpone a more detailed discussion to \sref{sec:washout}.  We focus on the Type-I seesaw model for concreteness, but our conclusions are immediately generalized to other Majorana neutrino mass models. We separate the discussion into two different mass regimes, $m_N \gg H_I$ and $m_N < H_I$.

\subsubsection{High Majorana mass scale, $m_N \gg H_I$}

Assuming that the additional heavy Majorana neutrinos are sufficiently massive compared to the inflationary Hubble scale, $m_N \gg H_I$, then they are not generated by a growing gravitational wave chirality during inflation.  Instead the lepton number is carried only by the standard model fermions, as we have discussed already in \sref{sec:BaryonAsym}.  This is the scenario proposed in Ref.\ \cite{Alexander:2004us}.  However, the lepton-number-violating Weinberg operator provides a channel to (partially) washout the lepton asymmetry.  

Typically the scale of explicit lepton-number violation (seesaw scale) is around $10^{12} - 10^{14} \GeV$, and we anticipate a significant washout of lepton number if the reheat temperature is as high as $T_\RH \sim 10^{15} \GeV$ as suggested by the estimates in Eq.\ \pref{eq:etaB_numerical} in the previous section. 
These processes have been ignored in existing studies of gravitational leptogenesis even though the assumption of instant reheating is often used in order to maximize the lepton asymmetry.  
In the next section we estimate the effects of these processes during reheating.  

\subsubsection{Low Majorana mass scale, $m_N \ll H_I$}

For $m_N \ll H_e$ the heavy Majorana neutrinos are produced gravitationally during inflation, and they carry a particle-antiparticle asymmetry given by $n_{\nu_R^i}(t_e) = H_e^3 \Ncal_{\B-\L}(t_e)/3$ at the end of inflation.  (We assume three roughly degenerate heavy neutrinos, $\nu_R^i$, but our conclusions are not qualitatively changed if one or two neutrinos are heavier and decoupled.)  Consequently, the $\B-\L$ asymmetry carried by the standard model species is cancelled, and there is no net baryon or lepton asymmetry.  Amusingly, this does not preclude the viability of baryogenesis.  This is because some of the asymmetry is carried by sequestered sectors.  {In particular, the asymmetry carried by right-chiral leptons can only be exchanged with other standard model particles through the respective Yukawa interactions \cite{Campbell:1992jd}, which remain out of equilibrium until $T \lesssim 3 \times 10^{11} \GeV$, $1 \times 10^9 \GeV$, and $8 \times 10^5 \GeV$ for the third, second, and first generation leptons, respectively.} This means that, as long as the lepton-number-violating interactions mediated by the $\nu_R^i$ go out of equilibrium before the lepton Yukawa interactions come into equilibrium, the lepton number carried by $e_R^2$ and/or $e_R^1$ can be transferred to the baryon asymmetry by the electroweak sphaleron.  We note that the condition $m_N \gg H_I$ was assumed by \rref{Alexander:2004us}. 

\section{Lepton-Number Washout in the Majorana mass scenario}
\label{sec:washout}

\begin{figure}[t]
\begin{center}
\includegraphics[width=0.45\textwidth]{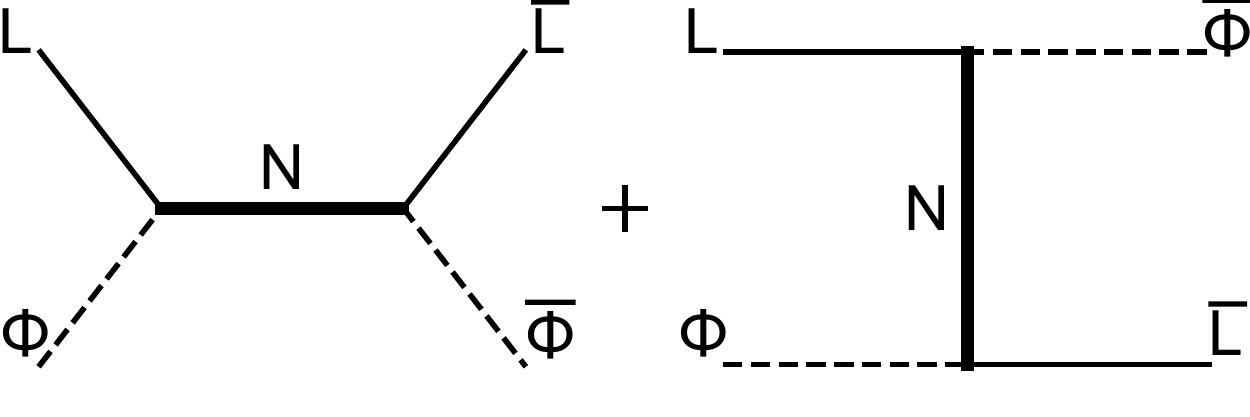} \\
\medskip
\includegraphics[width=0.45\textwidth]{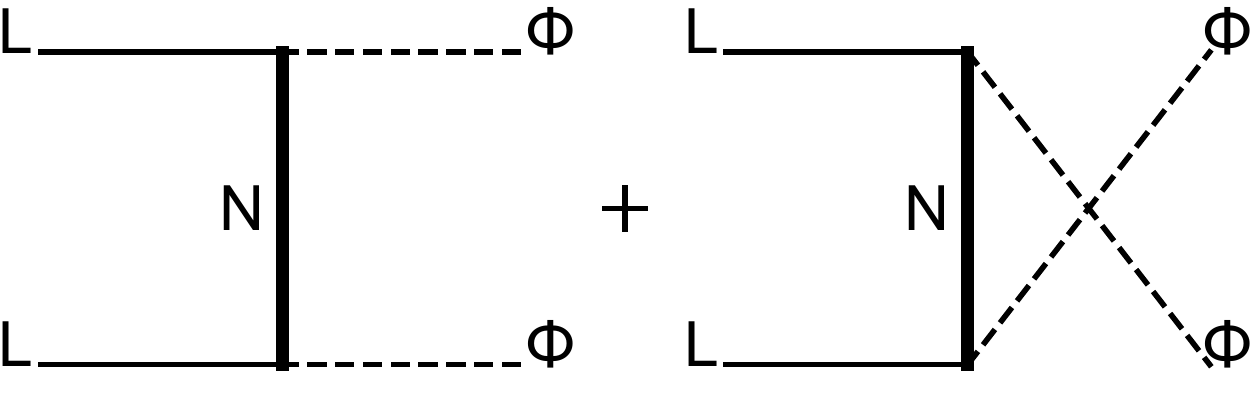}
\caption{\label{fig:washout}
Scattering processes in which a heavy Majorana neutrino $N$ mediates lepton-number-violating interactions among the standard model leptons $L$ and Higgs bosons $\Phi$. 
}
\end{center}
\end{figure}

The lepton-number washout calculation in this model is very similar to the standard analysis that one encounters in the study of thermal leptogenesis (see, {\it e.g.} Refs.~\cite{Giudice:2003jh,Strumia:2006qk} for detailed reviews).  For concreteness we assume that the light neutrino masses arise from the Type-I seesaw in which the standard model is extended to include three heavy right-chiral Majorana neutrinos, denoted by $N_i \equiv \nu_R^i$ for $i=1,2,3$, {with a common mass scale, $m_N$.} These heavy neutrinos mediate lepton-number-violating interactions among the standard model {left-chiral} leptons, denoted by $L_i = (\nu_L^i, e_L^i)$ for $i=1,2,3$, and the standard model Higgs bosons, denoted by $\Phi = (\phi^+, \phi^0)$.  These interactions are illustrated in \fref{fig:washout}.  

The thermally-averaged lepton-number washout rate can be calculated from the Feynman graphs in \fref{fig:washout} using the techniques described in Ref.~\cite{Giudice:2003jh}.  Here we make a rough estimate, which is reliable up to $O(1)$ numerical factors.  The thermal averaging consists of integrating over the energy of the external particles and weighting the cross section by the corresponding phase space distribution function.  For the $s$-channel process, the thermal averaging picks up a contribution from {energies with} $E \sim m_N$, where the intermediate right-handed neutrino ($N$) propagator goes on-shell. Otherwise the $N$ is off-shell, and for $T \ll m_N$ it is very off-shell.  At a time when the standard model plasma has a temperature $T$, the washout rate is estimated to be
\begin{align}\label{eq:Gam_wo}
	\Gamma_{\rm w.o.} \sim {\rm max}\Bigl[ 
	\frac{\lambda_N^2}{48\pi} \, \frac{m_N^3}{T^2} \, K_1\bigl( m_N / T \bigr)
	\ , \ 
	\frac{\lambda_N^4}{4\pi} \frac{T^3}{m_N^2} 
	\Bigr] 
	\per
\end{align}
The first term is the contribution from on-shell $N$'s, which is Boltzmann suppressed for $m_N \gg T$, and the second is the contribution from off-shell $N$'s.  Here $\lambda_N$ denotes the coupling associated with the $L\Phi N$ Yukawa interaction, $m_N$ is the mass of the heavy Majorana neutrinos $N_i$ (assumed to be approximately degenerate), and $K_n(x)$ is the modified Bessel function of the second kind of order $n$.  Since the $L\Phi N$ Yukawa interaction gives rise to the light neutrino masses after electroweak symmetry breaking, we can write $\lambda_N \approx \sqrt{2 m_N m_\nu / v^2}$ where $m_\nu \sim 10^{-10} \GeV$ is the light neutrino mass scale and $v \simeq 246 \GeV$ is the vacuum expectation value (VEV) of the Higgs field.   Note that once we specify $m_\nu$ and $v$, the off-shell contribution does not explicitly depend on the value of $m_N$.  We can see this result more directly by first integrating $N$ out of the theory to obtain the Weinberg operator, $(\lambda_N^2/m_N) L \Phi L \Phi$, and then calculating $\Gamma_{\rm w.o.}$, which corresponds to the second term in \eref{eq:Gam_wo}.  In this sense, the off-shell contribution to $\Gamma_{\rm w.o.}$ is ``model-independent'' and insensitive to the specific UV completion of the Weinberg operator.  

In this section we focus on the regime $m_N \gg H_e$ such that the heavy Majorana neutrinos are not produced gravitationally during inflation, and we discuss in \sref{sec:lower_mass} how are results are changed when this assumption is relaxed.  We have seen in \sref{sec:BaryonAsym} that gravitational leptogenesis favors large $H_e \sim 10^{13} \GeV$.  For such large values of $m_N$, the on-shell contribution to the washout rate in Eq.\ \pref{eq:Gam_wo} is negligible, and therefore we keep only the off-shell contribution in our numerical analysis.  

To determine the effect of washout on the baryon asymmetry, we solve the full system of standard model kinetic equations [see \rref{Kamada:2016eeb} for a summary], which are extended to include the collision terms corresponding to the additional lepton-number-violating interaction.  The new terms only appear in the kinetic equations for the left-chiral lepton asymmetries and the Higgs asymmetries; they are written as 
\begin{align}\label{eq:kin_eqns}
	d n_{\nu_L^i} / dt & \! \supset - \sum_{j=1}^{3} \Scal_{\nu h \nu h}^{ij} 
	\ , \ \ 
	d n_{\phi^0} / dt \supset - \sum_{i,j=1}^{3} S_{\nu h \nu h}^{ij} \nonumber \\ 
	d n_{e_L^i} / dt & \supset - \sum_{j=1}^{3} S_{e h e h}^{ij} 
	\ , \ \  
	d n_{\phi^+} / dt \supset - \sum_{i,j=1}^{3} \Scal_{e h e h}^{ij}
\end{align}
where
\begin{subequations}
\begin{align}
	S_{\nu h \nu h}^{ij} & = \Gamma_{\rm w.o.} ( n_{\nu_L^i} + n_{\nu_L^j} + n_{\phi^0}/2 + n_{\phi^0}/2 ) \, \delta^{ij} \\ 
	S_{e h e h}^{ij} & = \Gamma_{\rm w.o.} ( n_{e_L^i} + n_{e_L^j} + n_{\phi^+}/2 + n_{\phi^+}/2 ) \, \delta^{ij}
	\per 
\end{align}
\end{subequations}
For simplicity we assume that the lepton-number-violating interactions are flavor diagonal (in the same basis that diagonalizes Yukawa and gauge interactions) and flavor universal; hence, the Kronecker delta $\delta^{ij}$ appears.  The additional factors of $1/2$ on the Higgs terms are the result of Bose-Einstein statistics.  

Using the expressions above one can deduce that, while the washout processes are active, the baryon-minus-lepton-number density evolves according to the Boltzmann equation
\begin{align}\label{eq:Bmann_eqn}
\frac{d}{dt}n_{\B-\L} + 3H n_{\B-\L} = \Gamma_{\rm w.o.} \Bigl( 2 \sum_{i=1}^3 n_{l_L^i} + 3 n_{\Phi} \Bigr)
\end{align}
where $n_{l_L^i} = n_{\nu_L^i} + n_{e_L^i}$ and $n_{\Phi} = n_{\phi^+} + n_{\phi^0}$.  The source term from gravitational leptogenesis is absent after the end of inflation.  

\subsection{Semi-Analytical Solution}\label{sec:semianalytical}

Let us now derive a semi-analytical solution to the system of kinetic equations and \eref{eq:Bmann_eqn} in particular.  It is useful to first express the right side of \eref{eq:Bmann_eqn} in terms of $n_{\B-\L}$.  To do this we focus on plasma temperatures around $T \sim 10^{11} \GeV$, {corresponding roughly to lowest temperature at which the lepton-number-violating interactions are still in equilibrium ($\Gamma_{\rm w.o.} \sim H$)}. The  standard model processes that are in thermal equilibrium are the weak sphaleron, the strong sphaleron, and the third generation up-type quark Yukawa interaction (see \rref{Kamada:2016eeb} for further details).  The reactions that are out of equilibrium imply effective conservation laws.  Importantly, since the lepton Yukawa interactions are out of equilibrium, the corresponding right-chiral lepton-number densities, $n_{e_R^i}$, are effectively conserved \cite{Campbell:1992jd}.  Solving the resulting system of equilibrium conditions and conservations laws for $n_{l_L^i}$ and $n_\Phi$ lets us express the right side of \eref{eq:Bmann_eqn} as
\begin{align}\label{eq:Bmann_eqn_RHS}
	2 \sum_{i=1}^3 n_{l_L^i} + 3 n_{\Phi} = - \frac{348}{115} n_{\B-\L} + \frac{72}{115} (n_{e_R^1} + n_{e_R^2} + n_{e_R^3}) 
	\per 
\end{align}
The numerical coefficients are related to an accounting of the degrees of freedom and the hypercharge assignments.  

Now we understand how the solution of \eref{eq:Bmann_eqn} behaves.  The first term in \eref{eq:Bmann_eqn_RHS} tends to washout the initial baryon-minus-lepton asymmetry as long as $\Gamma_{\rm w.o.} > H$.  However, the second term prevents $n_{\B-\L}$ from dropping exponentially close to zero, instead $n_{\B-\L}$ saturates to a finite value, even when  the left-chiral lepton-number-violating interactions from \fref{fig:washout} are in equilibrium.  The lepton number carried by the right-chiral leptons, $e_R^i$, is protected from washout, because the charged lepton Yukawa interactions are out of equilibrium while the left-chiral lepton-number-violating interactions are in equilibrium.  

We can derive a semi-analytic solution to the Boltzmann equation above.  We first consider the regime where the second term in \eref{eq:Bmann_eqn_RHS} is negligible, and \eref{eq:Bmann_eqn} can be written as $dn_{\B-\L}/dt + 3 H n_{\B-\L} = - \Ccal \, \Gamma_{\rm w.o.} \, n_{\B-\L}$ where $\Ccal = 384/115 \simeq 3.03$.  Upon specifying the boundary condition at the end of inflation ($t=t_e, a=a_e$), the solution is 
\begin{align}\label{eq:nBmL_soln}
	{n_{\B-\L}(t) = n_{\B-\L}(t_e)} \left( \frac{a(t)}{a_e} \right)^{-3} \varepsilon_{\rm w.o.}(t)
\end{align}
where {the washout factor is} 
\begin{align}\label{eq:epsilon_wo}
\varepsilon_{\rm w.o.}(t) = {\rm exp}\left[ - \, \Ccal \int_{a_e}^{a(t)} \! \frac{\ud a^\prime}{a^\prime} \, \frac{\Gamma_{\rm w.o.}(T(a^\prime))}{H(a^\prime)} \right] \per
\end{align}
Note that $\varepsilon_{\rm w.o.}$ asymptotes to a constant at late times when $\Gamma_{\rm w.o.} \ll H$.  Therefore $\lim_{t \to \infty} \varepsilon_{\rm w.o.}(t)$ gives the suppression of the baryon-minus-lepton asymmetry due to washout.  
To further evaluate $\varepsilon_{\rm w.o.}$ it is necessary to select a model of reheating, which specifies $T(a)$ and $H(a)$, and we return to this point in \sref{sec:reheating}.

If lepton-number violation is very efficient, $\Gamma_{\rm w.o.} \gg H$, then \erefs{eq:nBmL_soln}{eq:epsilon_wo} imply an exponentially small value for $n_{\B-\L}$.  However, the second term in \eref{eq:Bmann_eqn_RHS} leads instead to a finite asymptotic value where $n_{\B-\L} = (6/29)( n_{e_R^1} + n_{e_R^2} + n_{e_R^3} )$.  Since the comoving densities of $e_R^i$ are conserved, we can relate these densities directly to the initial condition from gravitational leptogenesis, see \eref{eq:initial}.  Doing so gives $n_{\B-\L} = \varepsilon_{\rm w.o.} \, n_{\B-\L}^{(0)}$ where $n_{\B-\L}^{(0)} = H_e^3 \Ncal_{\B-\L}(t_e) \, (a(t) / a_e)^{-3}$ would be the value of $n_{\B-\L}$ if $(\B-\L)$ were conserved and there were no washout, and where $\varepsilon_{\rm w.o.} = 6/29 \simeq 0.21$.  {We demonstrate below in \sref{sec:numerical} that} this calculation matches well the fully numerical solution that appears in \fref{fig:PR_washout}.  

\subsection{Reheating}\label{sec:reheating}

At the end of inflation, the inflaton must transfer its energy into the standard model particles.  This is accomplished through either perturbative decay \cite{Abbott:1982hn,Albrecht:1982mp}, non-perturbative parametric resonance \cite{Traschen:1990sw, Shtanov:1994ce, Kofman:1994rk} (preheating), or possibly both mechanisms. For this work, we assume perturbative reheating and that the standard model sector thermalizes quickly forming a hot plasma.  This plasma does not cool adiabatically, because it continues to be populated by the inflaton decay products.  In fact, the standard model plasma reaches a maximum temperature $T_{\rm max} \approx T_\RH (H_e / H_\RH)^{1/4}$ \cite{Chung:1998rq,Giudice:2000ex}, and it cools as $T(a) \sim a^{-3/8}$ during the reheating epoch ($a_e < a < a_\RH$).  Meanwhile the total energy density is still dominated by the inflaton, which redshifts like pressureless dust $\rho \sim a^{-3}$, and consequently the Hubble scale evolves as $H \sim a^{-3/2}$, which corresponds to $w = 0$ in \eref{eq:expansion_factor}.  After reheating is completed ($a = a_\RH$) the energy density of the standard model plasma is dominant, implying $H \sim a^{-2}$, and the plasma cools adiabatically, implying $T \sim a^{-1}$.  

It is straightforward to phenomenologically modify this model of reheating to allow for different equations of state in order to examine the conditions in which the washout of lepton number can be minimized or avoided.    We take a phenomenological approach and simply use a more general equation of state for the inflaton $p=w\rho$, which allows for $w\ne 0$.  

For a general, constant $w$, the behavior of the temperature and Hubble scale during and after reheating are well approximated by  (generalizing the computation of \rref{Chung:1998rq} to $w \neq 0$)
\begin{subequations}
\label{eq:TH_pert_reheating}
\begin{align}
	T(a) & = \begin{cases} 
	T_{\rm max} \left( a / a_e \right)^{-3(1+w)\over 8} & \mathrm{for}\ a_e \leq a < a_{\RH} \\
	T_{\RH} \left( a / a_{\RH} \right)^{-1} & \mathrm{for}\ a_{\RH} \leq a 
	\end{cases} \\ 
	H(a) & = \begin{cases} 
	H_e \left( a / a_e \right)^{-3(1+w)\over 2} & \mathrm{for}\ a_e \leq a < a_{\RH} \\
	H_{\RH} \left( a / a_{\RH} \right)^{-2} & \mathrm{for}\ a_{\RH} \leq a 
	\end{cases} 
\end{align}
\end{subequations}
where 
\begin{align}\label{eq:Tmax_def}
	\left( \frac{a_{\RH}}{a_{e}} \right)^{3(1+w)} = \left( \frac{T_{\rm max}}{T_{\RH}} \right)^{8} = \left( \frac{H_{e}}{H_{\RH}} \right)^{2} 
	\per
\end{align}
The temperature and Hubble parameter at the end of reheating, $T_\RH$ and $H_\RH$, are determined by the inflaton decay rate $\Gamma_\phi$.  Approximately, the relation is $H_\RH \approx \Gamma_\phi$ or $T_\RH \sim \sqrt{\Gamma_\phi M_{\rm Pl}}$. In the following we treat $T_\RH$ as a free parameter.  The maximum  temperature during reheating, $T_{\rm max}$, (at fixed values of $H_e$ and $T_{\RH}$) does not change significantly with $w$ as compared to the usual $w=0$ case studied in Ref.\ \cite{Chung:1998rq}. We find that $T_{\rm max}$ is about $3\%$ larger for $w=-1/3$ and about $7\%$ smaller for the extreme case of $w=1$.

\subsection{Fully Numerical Solution}\label{sec:numerical}

\begin{figure}[t]
\begin{center}
\includegraphics[width=0.45\textwidth]{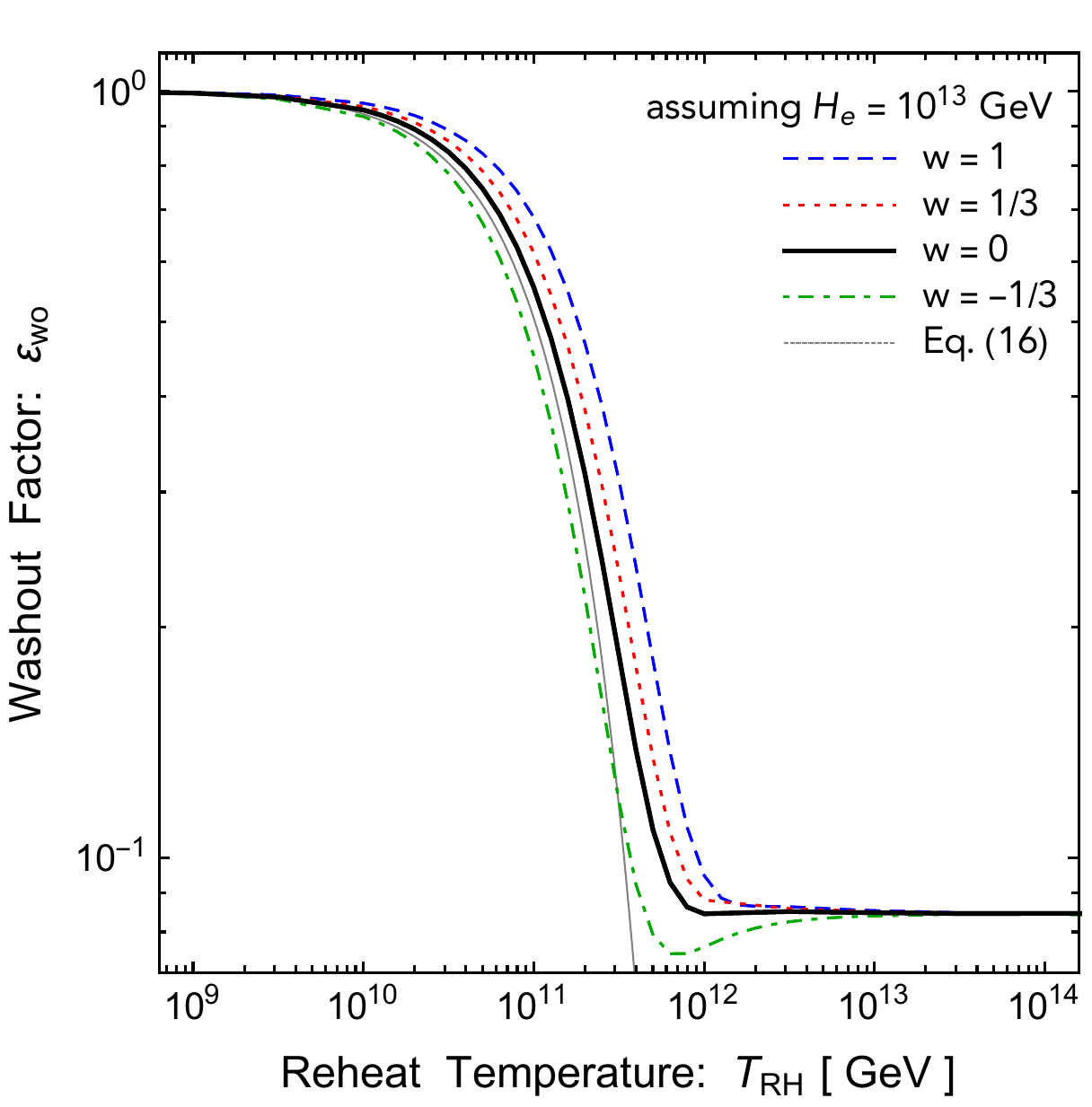} 
\caption{\label{fig:PR_washout}
We show the effective washout factor, $\varepsilon_{\rm w.o.}$, which corrects the formula in Eq.\ \eqref{eq:etaB_numerical} to account for lepton-number violation due to heavy right-handed neutrino exchange.  We vary the equation of state during reheating in the blue-dashed, red-dotted, black-solid, and green-dot-dashed lines.  The thin gray line shows the approximation in \eref{eq:epsilon_wo} for $w=0$.  At high reheat temperature $\varepsilon_{\rm w.o.} \simeq 0.09$, {but this value has an $O(1)$ uncertainty from our rough estimation of $\Gamma_{\rm w.o.}$ in \eref{eq:Gam_wo}}.  }
\end{center}
\end{figure}

\begin{figure*}[t]
\begin{center}
\includegraphics[width=0.45\textwidth]{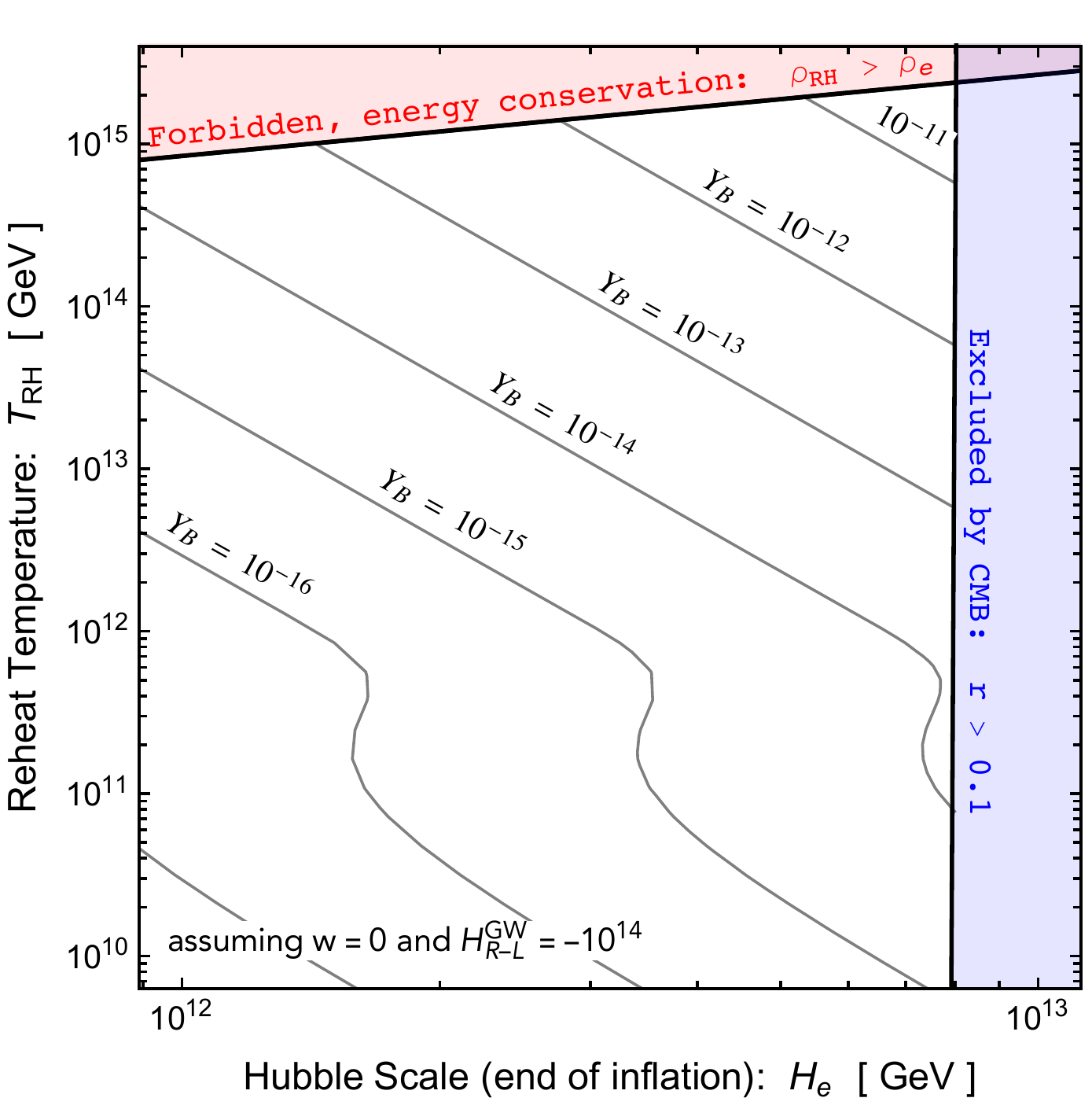}\hskip 1.5 cm \includegraphics[width=0.45\textwidth]{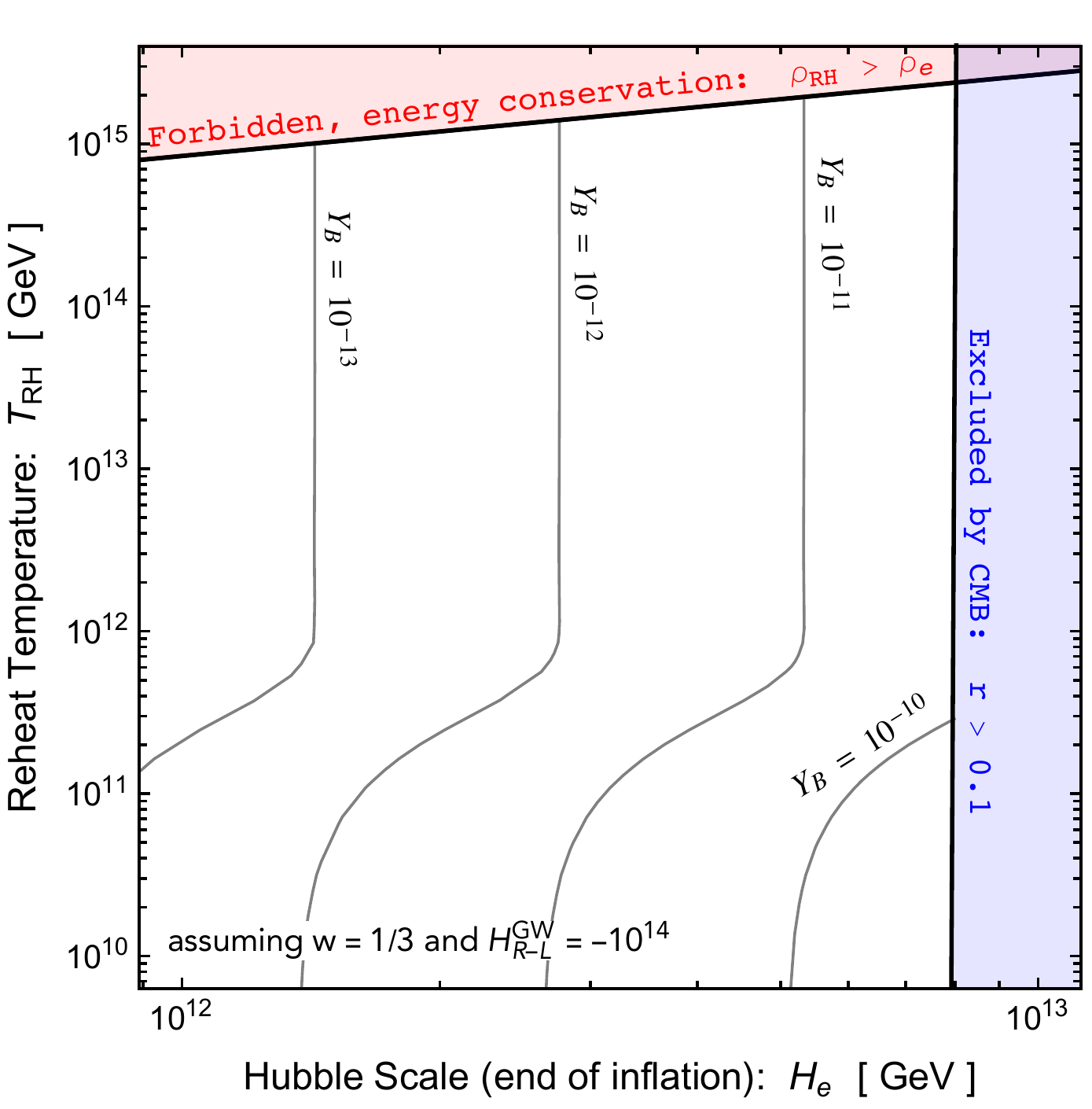}
\caption{\label{fig:param_space_2}
The baryon-to-entropy ratio $Y_\B = n_\B / s$ generated from gravitational leptogenesis in a Majorana-mass model where the Hubble scale at the end of inflation is $H_e$, the plasma temperature at the end of reheating is $T_\RH$, and the effective equation of state during reheating is $w = 0$ (left panel), and $w = 1/3$ (right panel).  We take $\Hcal_{R-L}^\GW = -10^{14}$ to draw the contours, but more generally $Y_\B \propto -\Hcal_{R-L}^\GW$.  The washout of lepton number by approximately an order of magnitude is apparent for $T_\RH \gtrsim 10^{11} \GeV$.  
}
\end{center}
\end{figure*}

Upon including washout effects, we numerically solve the full system of kinetic equations, {i.e. the equations in \rref{Kamada:2016eeb} extended by the terms in \eref{eq:kin_eqns},} to determine the baryon asymmetry $Y_\B$.  We define the washout suppression factor $\varepsilon_{\rm w.o.}$ as the ratio of this $Y_\B$ and the analytic formula for $Y_\B$ in \eref{eq:etaB_numerical}.  To compare with with the semi-analytical calculation, we also use \eref{eq:TH_pert_reheating} to evaluate the integral that appears in \eref{eq:epsilon_wo}.  The integral can be written in terms of special functions, but the resulting expression is not particularly illuminating, and we do not present it here.  Instead, we present the results of integrating \eref{eq:epsilon_wo} graphically in \fref{fig:PR_washout}.  We show the factor by which the net lepton number is washed out for four different expansion histories during the reheating phase parameterized by equations of state $w \in \{-1/3, 0, 1/3, 1\}$.  The washout factor is only weakly dependent upon $H_e$ for $10^{12} \GeV < H_e < 10^{13} \GeV$.  The fully numerical solution agrees very well with the semi-analytical solution that was derived in \sref{sec:semianalytical}.  

We now consider in detail the cases of matter dominated expansion during reheating, $w = 0$, and radiation dominated expansion during reheating, $w = 1/3$.

\subsubsection{Matter domination, $w = 0$}

If reheating after inflation proceeds via the perturbative decay of a massive inflaton oscillating about the minima of quadratic potential, the equation of state during reheating is very close to that of dust, $w = 0$ \cite{Turner:1983he}.  We present our results numerically for this case in the left panel of  \fref{fig:param_space_2}, which shows the dependence of the final baryon asymmetry $Y_\B$ on the reheat temperature $T_\RH$ and the Hubble rate at the end of inflation.  The predicted $Y_\B$ is insensitive to the mass scale of the heavy Majorana neutrinos provided that $m_N$ is large enough for the off-shell contribution to $\Gamma_{\rm w.o.}$ to dominate [second term in \eref{eq:Gam_wo}].  We discuss the regime with smaller $m_N$ in \sref{sec:lower_mass}.  
Note that as the reheating temperature drops below $T_\RH \simeq 1 \times 10^{11} \GeV$ washout becomes negligible and the iso-baryon asymmetry curves approach the curves in \fref{fig:param_space_1}. 

The requirements for successful gravitational leptogenesis can be read-off by simply looking at Fig.~\ref{fig:param_space_1} and Fig.~\ref{fig:PR_washout}.  Neglecting any washout processes, observationally viable values of $Y_B$ are generated for large values of the Hubble scale $H\gtrsim 5\times 10^{12}\, {\rm GeV}$ and also large values of the reheat temperature $T_{\RH} \gtrsim 10^{14}\, {\rm GeV}$, close to the instantaneous reheating limit.  Dilution of the lepton-number density due to the expansion of the universe during a matter-dominated phase makes this corner of parameter space the only viable one. Since for large values of the reheat temperature the washout factor is constant $\epsilon_{w.o.}\sim 0.08$, successful gravitational leptogenesis requires increasing the initial asymmetry by about one or two orders of magnitude, corresponding to $|\Hcal_{R-L}^\GW| \gtrsim 10^{15}$, in order to counteract the washout, while keeping the same high values of the Hubble scale and reheat temperature. 

\subsubsection{Radiation domination, $w = 1/3$}

In Chromo-Natural inflation, or Gauge-flation, the universe is dominated by a very weakly coupled  ($g \lesssim 10^{-5}$) gauge field at the end of inflation. In these cases, the universe transitions quickly {(within $\sim 3$ $e$-folds)} to expanding with an effective equation of state of $w=1/3$, corresponding to radiation domination. Reheating in this case is facilitated by the decay of the (dark) gauge bosons into the standard model. Similar behavior could also arise, for example, if the inflaton decays exclusively into a dark sector with $w=1/3$, which then decays into the SM. In both of these cases, the standard model is not thermalized until some later time, denoted by $T_{\rm RH}$. The equation of state $w=1/3$ is also attained for a quartic potential, \cite{Turner:1983he, DeCross:2015uza}, and more generally for potentials that are different from quadratic at the origin \cite{Lozanov:2016hid, Lozanov:2017hjm}.  We present the numerical results for this case in the right hand panel of \fref{fig:param_space_2}. We observe similar behavior to the matter dominated case, namely that as the reheating temperature drops below $T_\RH \simeq 1 \times 10^{11} \GeV$ washout becomes negligible and the iso-baryon asymmetry curves approach the curves in \fref{fig:param_space_1}. Contrary to the matter-dominated reheating case, successful gravitational leptogenesis is possible for $|\Hcal_{R-L}^\GW| \sim 10^{14}$, which is within the realm of the modified CNI models considered by Ref.\ \cite{Caldwell:2017chz}.

\subsection{Lower Majorana Neutrino Mass}\label{sec:lower_mass}

In the preceding discussion we have assumed that the scale of the heavy Majorana neutrinos obeys $m_N \gg H_e$ such that these particles are not produced during inflation, and they do not thermalize with the standard model plasma.  {Then $\Gamma_{\rm w.o.}$ can be approximated by the off-shell contribution alone, which is the second term in \eref{eq:Gam_wo}.}  In this section we discuss how the previous results are changed when $m_N$ is lower.  

\begin{figure}[t]
\includegraphics[width=0.45\textwidth]{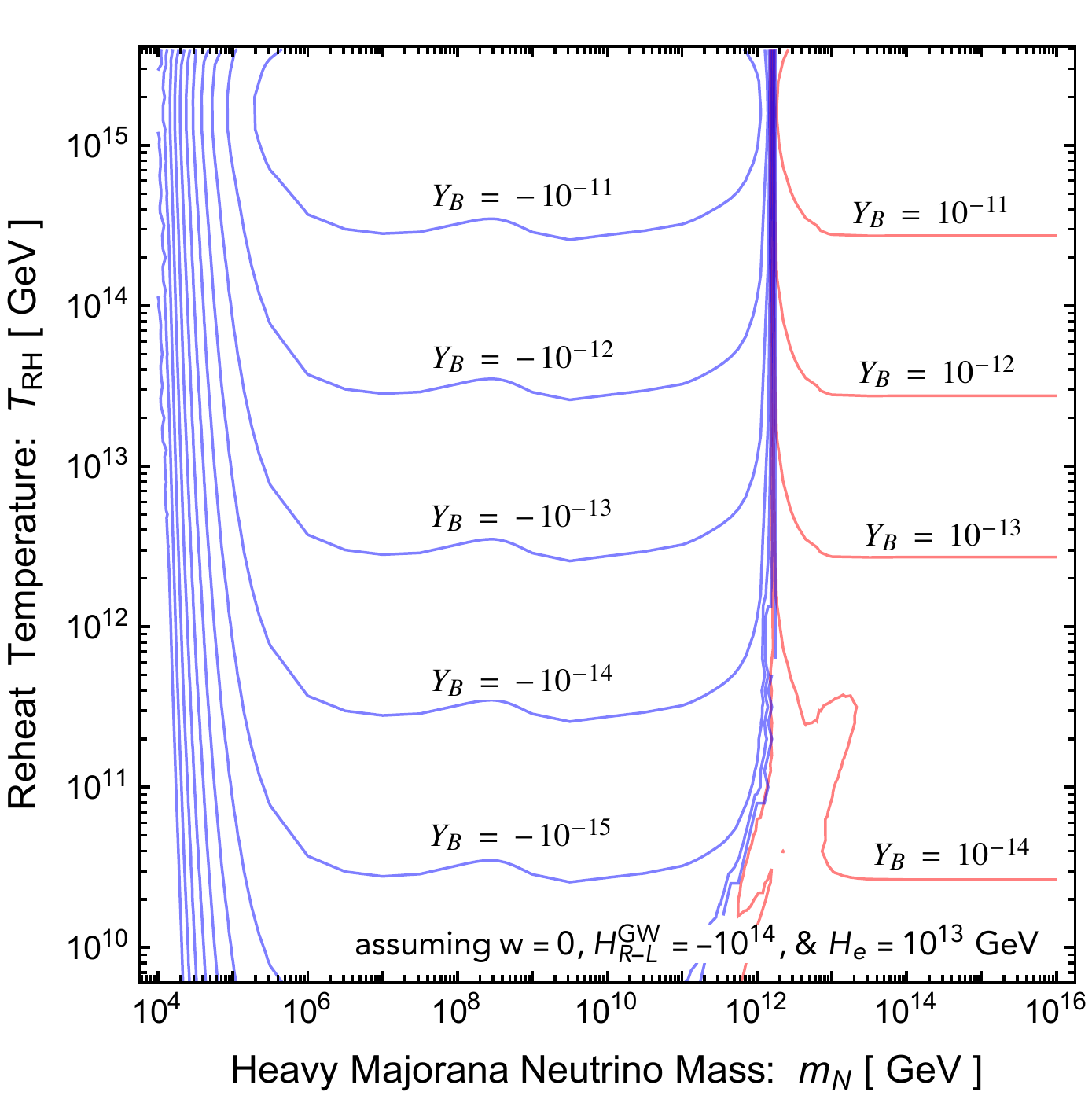} 
\caption{\label{fig:varym_N}
The effect of varying the mass scale of the heavy Majorana neutrinos, $m_N$, and the reheating temperature (for matter-dominated expansion, $w = 0$, during reheating) on the resulting baryon-to-entropy ratio $Y_\B = n_\B / s$ that is generated from gravitational leptogenesis. In making this figure, we have taken $H_e = 10^{13} \GeV$. Note that the baryon asymmetry changes sign at $m_{N}\sim 10^{12}$ GeV. For $m_{N} > 10^{12}$ GeV, sign($Y_{B}$) = $-$sign($\Hcal_{R-L}^\GW$), while  sign($Y_{B}$) = sign($\Hcal_{R-L}^\GW$) for $10^6 < m_{N} < 10^{12}$ GeV.}
\end{figure}

In Figure \ref{fig:varym_N} we numerically study the regime $m_N \ll H_e$ by including both the on-shell and off-shell contributions to the thermally averaged washout rate in \eref{eq:Gam_wo}.  We consider matter-dominated expansion during reheating, which dilutes the lepton asymmetry before reheating.  As demonstrated above, this dilution can be avoided if the equation of state is that of radiation.  For $m_N > 10^{13} \GeV$ the results are unchanged from the calculation in the previous section where the heavy Majorana neutrinos are decoupled.  For $m_N < 10^{13} \GeV$ the relic baryon asymmetry is modified by an $O(1)$ factor, and the sign flips.  This is because the lepton asymmetry carried by the left-chiral leptons is efficiently washed out, and the lepton asymmetry carried by the $e_R^i$ is eventually redistributed when the corresponding Yukawa interaction comes into equilibrium.  An exponential washout of the baryon and lepton asymmetries is avoided unless the heavy Majorana mass scale is very low, $m_N \lesssim 10^6 \GeV$, such that lepton-number violation is still in equilibrium when the electron Yukawa equilibrium comes into equilibrium (and $e_R^1$-conservation is lost).  

\section{Conclusions}\label{sec:conclusions}

In this work we have examined inflationary gravitational leptogenesis when confronted with realistic models of reheating and neutrino mass generation.  
Whereas it is customary to assume instantaneous reheating in studies of gravitational leptogenesis, models of reheating generally predict a much smaller reheat temperature, $T_\RH$. We study  the dependence of the predicted baryon asymmetry on $T_\RH$ and the effective equation of state during reheating, $w$.  Additionally, earlier studies of gravitational leptogenesis neglect the possible effects of nonzero neutrino mass, which requires new particles and interactions beyond the standard model.  In this work, we have studied the implications of both the Dirac and Majorana mass scenarios. We have shown that gravitational leptogenesis is viable in both mass scenarios, despite the fact that lepton number is not violated in the Dirac scenario, and despite the fact that the lepton asymmetry can be washed out in the Majorana scenario.  
In the remainder of this section, we summarize our key findings related to gravitational leptogenesis in the context of realistic models of reheating and neutrino mass generation.  

Relaxing the assumption of instantaneous reheating, we apply a phenomenological description of reheating to calculate the baryon asymmetry, $Y_\B$, in terms of the reheat temperature, $T_\RH$, and the equation of state during reheating, $w$  (assumed to  be constant).  Under these generalized assumptions, \eref{eq:etaB_numerical} gives the prediction for $Y_\B = n_\B / s$ which is illustrated in \fref{fig:param_space_1}.  If the universe is effectively matter dominated during reheating, $w = 0$, the baryon asymmetry is diluted, because the comoving number density, $a^3 n_{\B-\L}$, is conserved.  To avoid diluting $Y_\B$ excessively, the reheat temperature must be high, $T_\RH \gtrsim 10^{14} \GeV$ for the benchmark gravitational wave chirality assumed here, $\Hcal_{L-R}^\GW = - 10^{14}$; the limit weakens for larger $\Hcal_{L-R}^\GW$.  However, if the universe is radiation dominated during reheating, $w = 1/3$,  then the dilution factor is compensated by the $T_\RH$-dependence in the entropy density, $s$, and the resulting baryon asymmetry, $Y_\B = n_\B/s$, is independent of $T_\RH$.  In either scenario, gravitational leptogenesis requires a high Hubble scale at the end of inflation, $H_e \gtrsim 10^{12} \GeV$, which implies an amplitude of primordial gravitational waves that is within reach of CMB polarization (B-mode) measurements.  In a (more exotic) model with $w > 1/3$ the baryon asymmetry {\it increases} during reheating, and $Y_\B$ can be compatible with the measured asymmetry for a smaller gravitational wave chirality.  

Going beyond the standard model of particle physics, we first consider that the neutrinos are Dirac particles which get their tiny mass from a small Yukawa coupling.   Upon introducing three right-chiral neutrino fields, in order to fill out the missing components of the neutrino Dirac spinor, the gravitational anomaly in lepton number is vanishing, because the contributions from left- and right-chiral leptons cancel.  Nevertheless, gravitational leptogenesis is still a viable explanation of the matter-antimatter asymmetry.  Although the growing gravitational wave chirality does not generate a {\it net} lepton number, it does generate equal and opposite lepton asymmetries in the active and sterile neutrinos.  Since the neutrino Yukawa coupling is extremely tiny, the interactions it mediates are out of equilibrium, and the lepton number carried by the sterile neutrinos is effectively sequestered from the lepton number in the standard model sector.  Consequently, the predictions of gravitational leptogenesis are unaffected by the presence of the sterile neutrinos, and the resultant baryon asymmetry appears in \eref{eq:etaB_numerical} and \fref{fig:param_space_1}.   The relic sterile neutrinos are unobservable in practice.  

Finally we study gravitational leptogenesis under the hypothesis that the light neutrinos are Majorana particles, and the neutrino mass scale is set by the Type-I seesaw mechanism upon introducing much heavier right-chiral Majorana neutrinos.  The heavy Majorana neutrinos mediate interactions that violate $(\B-\L)$ and threaten to wash out the $\B-\L$ asymmetry generated by gravitational leptogenesis.  However, we have shown that a complete (exponential) erasure of the asymmetry is avoided as long as the lepton Yukawa-interactions are out of equilibrium at the temperatures where the $(\B-\L)$ violation is in equilibrium.  This is because gravitational leptogenesis populates an asymmetry in all of the standard model fermions, and the lepton number carried by the right-chiral charged leptons is protected from washout while the lepton Yukawa-interactions are out of equilibrium.  Using both semi-analytical arguments and a fully numerical calculation, we show that the washout factor varies from $\epsilon_{\rm w.o.} \approx 1$ for $T_\RH \lesssim 10^{11} \GeV$ to a modest suppression of $\varepsilon_{\rm w.o.} \simeq 0.08$ for a higher reheat temperature (see  \fref{fig:PR_washout}).  

This sequestration of lepton number in right-chiral charged leptons also implies that the mass scale of the heavy Majorana neutrinos need not satisfy $m_N \gg H_I$, which is often assumed in studies of gravitational leptogenesis.  For $m_N < H_I$ both the left- and right-chiral neutrinos are populated during inflation, and the net lepton asymmetry vanishes, as in the Dirac mass scenario discussed above.  Subsequently, for a high enough reheat temperature, the asymmetries carried by the right-chiral Majorana neutrinos and the left-chiral leptons will be partially washed out by their $(\B-\L)$-violating Yukawa interactions.  However, as shown in \fref{fig:varym_N}, as long as $m_N \gtrsim 10^6 \GeV$, the $(\B-\L)$-violating interactions go out of equilibrium before the electron Yukawa interaction comes into equilibrium, and the lepton number carried by the right-chiral electron is preserved and converted to baryon number by the standard model electroweak sphaleron.  

Finally, we note that there is a definite connection between the chirality of the gravitational wave background, the nature of the neutrinos (Dirac vs Majorana), and the scale of explicit lepton-number violation (if present).   On the one hand, to explain the baryon asymmetry, left-chiral gravitational waves require either Dirac neutrinos \emph{or} Majorana neutrinos with high-scale lepton-number violation, $m_{N} \gtrsim 10^{12} \GeV$.   On the other hand, right-chiral gravitational waves require Majorana neutrinos with lepton number violation $10^{6} < m_{N} < 10^{12} \GeV$, which can be seen from \fref{fig:varym_N}.  Therefore, detection of either a right- or left-chiral gravitational wave background may shed light on the nature of neutrino mass generation.  

\acknowledgments
We thank Robert Caldwell, Daniel Chung, Lisa Everett, Yuta Hamada, Sonia Paban, Lauren Pearce, and Marco Peloso for useful conversations. The work of PA and EIS was supported in part by NASA Astrophysics Theory Grant NNX17AG48G. This work was initiated at the Aspen Center for Physics, which is supported by National Science Foundation grant PHY-1607611.  A.J.L. is supported at the University of Chicago by the Kavli Institute for Cosmological Physics through grant NSF PHY-1125897 and an endowment from the Kavli Foundation and its founder Fred Kavli. EIS gratefully acknowledges support from a Fortner Fellowship at the University of Illinois at Urbana-Champaign, and also the Dutch Organisation for Scientific Research (NWO).

\bibliography{Gravilepto}
\end{document}